\documentstyle[11pt,aaspp4,psfig]{article}
\def\be{\begin{equation}}
\def\ee{\end{equation}}

\def\lsim{\lower 2pt \hbox{$\, \buildrel {\scriptstyle <}\over
         {\scriptstyle \sim}\,$}}
\newcommand\gsim{\buildrel > \over \sim}
\begin{document}
\newcommand{\figureout}[3]{\psfig{figure=#1,width=8in,angle=#2} 
   \figcaption{#3} }

\title{Regimes of Pulsar Pair Formation and Particle Energetics}

\author{Alice K. Harding\altaffilmark{1}, Alexander G. Muslimov\altaffilmark{2} 
\& Bing Zhang\altaffilmark{3}
}
\altaffiltext{1}{Laboratory of High Energy Astrophysics, 
NASA/Goddard Space Flight Center, Greenbelt, MD 20771}

\altaffiltext{2}{ManTech International Corp., Lexington Park, MD 20653}

\altaffiltext{3}{Astronomy \& Astrophysics Department, Pennsylvania State University, 
Pennsylvania, PA 16802}  
%\slugcomment{To appear in The Astrophysical Journal, *** issue.}

\begin{abstract}

We investigate the conditions required for the production of electron-positron 
pairs above a pulsar polar cap (PC) and the influence of pair production on the energetics 
of the primary particle acceleration.  Assuming space-charge limited flow acceleration 
including the inertial frame-dragging effect, we allow both one-photon and two-photon pair 
production by either curvature radiation (CR) photons or photons resulting from inverse-Compton
scattering of thermal photons from the PC by primary electrons.  We find that, while
only the younger pulsars can produce pairs through CR, 
nearly all known radio pulsars are capable of producing pairs through non-resonant
inverse-Compton scatterings.  The effect of the neutron star equations of state on the
pair death lines is explored.  We show that pair production is facilitated in more 
compact stars and more massive stars.  Therefore accretion of mass by pulsars in
binary systems may allow pair production in most of the millisecond pulsar population.
We also find that two-photon pair production may be important in millisecond pulsars
if their surface temperatures are above $\simeq $ three million degrees K.  Pulsars that
produce pairs through CR will have their primary acceleration limited 
by the effect of screening of the electric field.  In this regime, the high-energy luminosity should
follow a $L_{HE} \propto \dot E_{rot}^{1/2}$ dependence.  The acceleration voltage drop in 
pulsars that produce pairs only through inverse-Compton emission will not be limited
by electric field screening.  In this regime, the high-energy luminosity should
follow a $L_{HE} \propto \dot E_{rot}$ dependence.  Thus, older pulsars will have
significantly lower $\gamma$-ray luminosity.

\end{abstract} 


\pagebreak
%\received{}
%\revised{}
%\accepted{}
  
\section{INTRODUCTION}

The acceleration of particles and the production of electron-positron pairs 
are widely considered to be two critical elements necessary for generating 
radiation in rotation-powered pulsars.  In polar cap models (e.g. Arons \& Sharlemann 1979, Daugherty \& Harding 1996), acceleration occurs in the region 
of open field near the magnetic poles and $\gamma$-rays from curvature and inverse Compton radiation produce pairs primarily by one-photon
pair production in the strong magnetic field.  These pairs may screen the 
accelerating electric field through the trapping and reversal of one sign of charge, 
and may be required for the coherent radio emission process.  In outer
gap accelerators (e.g. Cheng, Ho \& Ruderman 1986), a vacuum gap develops along 
the null charge surface and pairs are required to provide current flow through 
the gap which can then operate as a stable accelerator.  

In this paper we discuss plausible regimes of pair formation above the pulsar 
polar cap (PC), including the energetics of relativistic particles and 
$\gamma $-rays that cause and accompany these regimes. We treat the 
acceleration of particles within the framework of an approach elaborated by 
Harding \& Muslimov (1998, hereafter HM98), Harding \& Muslimov (2001, 
hereafter HM01), and Harding \& Muslimov (2002, hereafter HM02) which combines 
rough analytic estimates and simple practical formulae with detailed numerical 
calculations. As an underlying PC acceleration model, we employ the 
general-relativistic version of a space-charge limited flow model developed 
earlier by Muslimov \& Tsygan (1992, hereafter MT92) and advanced in a number 
of important aspects by HM98, HM01, and HM02. The main focus of our present study 
is the physical condition for pair formation and how this condition translates 
into a theoretical pair death line for the observed radio pulsars. This paper is a 
logical epilogue of our previous studies (see HM01 and HM02) where we calculated the 
parameters of pair-formation fronts (PFF), including the flux of returning positrons, 
calculated X-ray luminosities due to PC heating, estimated luminosity of the primary beam, 
and revised derivation of pulsar death lines. In our calculations of PFFs we employed 
the standard mechanism of magnetic pair-production by high-energy photons, generated via 
curvature radiation (CR) and/or inverse Compton scattering (ICS). In HM02 we calculated, both 
analytically and numerically, the theoretical pair death lines based on the abovementioned  
regimes of pair formation. However, in HM02 we presented the results of our calculation 
of pulsar death lines only for a canonical neutron star (NS) with the mass 1.4~$M_{\sun }$ 
and radius 10 km, even though we pointed out that the effect of deviation of NS mass and 
radius from their canonical values might be important for our calculations.   

In the present study we extend our previous analysis (HM02) of pulsar death lines to 
explicitly incorporate the effect of different NS mass and radius and, for the short-period 
(millisecond) pulsars, to include the possibility of two-photon pair formation. We 
must emphasize that the effect of bigger NS mass is especially worth considering for the 
millisecond (ms) pulsars, which are believed to be descendants of accreting NSs in low-mass binary 
systems. It is remarkable, that our present study suggests that the ms pulsars 
do favour the bigger NS masses which seems to be consistent with their standard evolutionary 
scenario. This effect is associated with the dominance of the relativistic frame-dragging 
term in the accelerating voltage drop which is a unique feature of the electrodynamic model 
of MT92. The frame-dragging component of the electric potential (field) is proportional to 
the general-relativistic parameter $\kappa = \epsilon I/MR^2$, where $\epsilon = r_g/R$, 
$r_g$ is the gravitational radius of a NS of mass $M$, and $I$ and $R$ are the stellar 
moment of inertia and mass, respectively.
Another important aspect of our previous and present studies is the derivation of 
a theoretical relationship between the pulsar's $\gamma$-ray luminosity and its  
spin-down power/luminosity. In this paper we discuss such a relationship in the context of 
the available and forthcoming $\gamma $-ray observations of pulsars.

The paper is organized as follows. In \S 2 we discuss the determination of pair death lines in pulsars. 
First, we outline the basic definition and main assumptions behind the death-line concept (\S 2.1). 
Second, we discuss the revised analytic approach in the derivation of theoretical death lines 
(\S 2.2), and then, in \S 2.3, we discuss our numerical calculation of death lines. In \S 2.3.1 
we calculate the CR and ICS death lines for the NS models with a canonical mass $1.4~M_{\sun }$ and 
with three different equations of state to illustrate the effect of compactness on our death line 
calculations. In \S\S 2.3.2, 2.3.3 we focus on the death line calculations for the short-period 
(millisecond) pulsars: we discuss how the change in mass and radius of the underlying NS model 
affects the pulsar death lines, and present the results of our numerical death line calculation 
(\S 2.3.2); we incorporate the effect of two-photon pair production for the PC temperatures 
1-5$\times 10^6$ K and present the corresponding numerical death lines (\S 2.3.3). In \S 2.3.2 
and 2.3.3 we illustrate separately how the mass of a NS and two-photon pair production, 
respectively, may affect the death lines for ms pulsars. In \S 3 we discuss energetics 
of the acceleration of primary electrons and present the theoretical relationship between the 
$\gamma $-ray luminosity and spin-down power of a pulsar. Finally, in \S 4 we summarize our 
main results and discuss their most important implications for pulsars.

\section{Death line determination in PSRs}

\subsection{General overview and definitions}

Since the very early attempts to relate the apparent absence of radio pulsars  
with long periods in the $P$--$\dot P$ diagram with the 
manifestation of the effect of electron-positron pair formation as 
a condition for their operation, it proved instructive to introduce 
the term 'death line' to separate the domain favouring pair formation 
from the domain where it would be prohibited (see Sturrock 1971; 
Ruderman \& Sutherland 1975; Chen \& Ruderman 1993, hereafter CR93; 
and references quoted therein). Soon, it became almost a common practice 
for any theoretical study on radio pulsars to produce the resulting 
death lines. Furthermore, some theories developed the idea that 
on the $P$-$\dot P$ map it is a death valley (see e.g. CR93) rather 
than a death line that separates radio active from radio quiet pulsars. 
It is important that during the past decade the number of new radio 
pulsars with a wide range of parameters dramatically increased, which 
boosted the various pulsar population studies. In light of the 
recent extensive radio pulsar surveys (e.g. Manchester et al. 2001), 
pulsar population studies, 
and multi-frequency (from radio to $\gamma $-ray) pulsar observations, 
it seems timely to get back to the basic concept of a pulsar death line.

The standard definition of a pulsar death line implies that pulsar radio 
emission turns off if the energetics of accelerated particles drops below  
the minimum required for electron-positron pair production. 
Thus, the standard definition of a death line implicitly assumes that 
pair-formation is a necessary condition for pulsar radio emission, and that 
pulsars become radio quiet after crossing the death lines during their 
evolution from left to right in the $P$-$\dot P$ diagram. Obviously, this 
basic condition may not be sufficient (see e.g. Hibschman \& Arons 2001, 
hereafter HA01; for a most recent study where the sufficient condition 
assumed was that of pair production with high enough multiplicity to compeletely
screen the parallel electric field), and 
functioning of radio pulsars may imply far more complex physical conditions
(see Usov 2002 for a recent review). 
However, numerous theoretical attempts to produce satisfactory death lines 
implying even the basic necessary condition meet certain challenges. For 
example, a number of observed ms radio pulsars fall below most 
theoretical death lines. Also, many normal radio pulsars tend to be below 
their death lines based on CR pair-formation. For this reason, and to 
minimize the underlying model assumptions, our previous (see HM02) and present 
analyses of pulsar death lines are based on the minimal requirement regarding  
pair formation. Note that in all our studies we assume a centered-dipole 
magnetic field of a NS. 

In this paper we illustrate how the spread of NS masses and radii may 
affect the theoretical death lines for ordinary and, most importantly, 
for ms radio pulsars. Needless to say, compactness of a NS 
is an important parameter in our calculations of particle acceleration 
(mostly because the accelerating electric field is of essentially 
general-relativistic origin) and pair formation (because of a bigger 
deflection of photon trajectories in the gravitational field of a more 
compact NS). So, the detailed analysis of the effects of stellar 
compactness on the results of such calculations would be quite 
interesting by itself. However, this effect is worthy of special 
consideration in the case of ms pulsars. The main reason 
is that the latter are believed to descend from accreting 
NSs in low-mass binary systems and may represent post-accreting 
relatively massive NSs. As will be demonstrated in Section 2.3.2, 
the increase in NS mass considerably facilitates pair formation 
in short-period pulsars, thus pushing the corresponding death lines 
down to or below the observed ($P$,$\dot P$) values for ms 
pulsars.

We also consider the effect of two-photon pair production, where $\gamma$-ray ICS photons 
interact with thermal X-ray photons from the NS surface to produce an electron-positron
pair.  This process is of primary importance in outer-gap models (e.g. Romani 1996; Zhang \& 
Cheng 1997) for 
pulsar high-energy emission.  Zhang \& Qiao (1998) investigated the importance of this 
process above PCs of normal pulsars, noting that unrealistically high PC temperatures were 
required.  As we will show in this paper, two-photon pairs can be produced more easily 
above PCs of ms pulsars because the PC size is larger. We will compute the death line for 
two-photon pairs as a function of PC temperature.

\subsection{Analytic death lines}

In this Section we generalize the analytic expressions for the death 
lines derived by Harding \& Muslimov (2002; hereafter HM02) to explicitly 
include the effect of different mass, radius, and moment of inertia 
of a NS. 

To formulate the analytic death line condition we need to know the 
distribution of voltage drop in the pulsar's PC acceleration region. 
For a given distribution of voltage we can calculate the characteristic 
Lorentz factor of a primary electron accelerating above the PC as a function 
of altitude $z$ and pulsar parameters $B$ and $P$, $\gamma _{acc}(z,B,P)$. 
The accelerated electron generates (CR and/or ICS) photons that 
may pair produce if the condition for the corresponding pair-formation 
process (in most cases magnetic pair production) is satisfied. In our 
previous papers (see e.g. HM98, HM01, and HM02) we have demonstrated that 
use of the pair-formation condition allows us to determine the pair-formation 
altitude as a function of pulsar parameters $B$ and $P$ alone. Then, the Lorentz 
factor of a primary electron evaluated at the pair-formation altitude determines 
a minimum Lorentz-factor an electron should achieve to generate a pair-producing 
photon, $\gamma _{\rm min}(B,P)$. Thus, the death line condition would require 
that the Lorentz-factor of an accelerating primary electron is equal to 
$\gamma _{\rm min}$. In our numerical calculation of death lines we can easily 
keep track of the fulfillment of this requirement and plot the corresponding points 
in the $P$--$\dot P$ diagram that constitute the death line (or rather death curve). 
However, an analytic derivation of the death line needs  an additional 
independent relationship between the pair-formation altitude $z$ and 
pulsar parameters $B$ (or $\dot P$) and $P$. In our previous paper (HM02) we 
have demonstrated that $\gamma _{acc}$ can be expressed as a function 
of $B$ , $P$ and an additional parameter, the efficiency of converting  
pulsar spin-down power into the luminosity of the primary beam, 
\be
f_{\rm prim} = L_{\rm prim}/{\dot E}_{\rm rot}, 
\label{fprim}
\ee
where $L_{\rm prim}$ is the luminosity of the primary electron beam, 
and ${\dot E}_{\rm rot}$ is the pulsar spin-down power 
($= \Omega ^4B_0^2R^6/6c^3f(1)^2$, where $B_0/f(1)$ 
is the surface value of the magnetic field strength corrected for the general-
relativistic red shift; all other quantities have their usual meaning and will 
be defined below; see also HM02 for details).  

For the typical radio pulsar parameters $P$ and $\dot P$ and for most 
obliquities, excluding the pure orthogonal case, the dominant term in the 
expressions for the electrostatic potential and electric field in the 
general-relativistic version of the space-charge limited flow model (MT92) 
is proportional to parameter $\kappa $. In this paper we shall use the 
following general expression for the parameter $\kappa $ to include its 
explicit dependence on NS radius and moment of inertia 
\be
\kappa \equiv \epsilon I/MR^2 = 0.15 \times I_{45}/R_6^3,
\label{kappa}
\ee
where $\epsilon $ is a NS compactness parameter, $I_{45}= I/10^{45}$ g$\cdot $cm$^2$,  
$R_6 = R/10^6$ cm; $M$, $R$ and $I$ are NS mass, radius and moment of inertia, 
respectively. 

Then, the explicit expression for $\gamma _{\rm acc}$ can be written as 
(cf HM02, equation [51])
\be
\gamma _{\rm acc} = 10^7 f_{\rm prim} B_{12} P^{-2},
\label{gamma_acc}
\ee
where
\be
f_{\rm prim} = {\bar f}_{\rm prim} ~\kappa _{0.15}~R_6^{-5/2}.
\label{fprim2}
\ee
and the $z$ dependence of $\gamma _{\rm acc}$ in implicit in $f_{\rm prim}$. 
Here ${\bar f}_{\rm prim}$ is the efficiency of converting spin-down 
luminosity into the luminosity of the primary beam, calculated 
for a canonical NS mass $M = 1.4~M_{\sun }$, radius $R = 10^6$ cm, and moment of 
inertia $I = 10^{45}$ g$\cdot $cm$^2$; and $\kappa _{0.15} = \kappa /0.15$; 
$B_{12} = B_0/10^{12}$ G, $B_0$ is the surface value of the magnetic field 
strength; $P$ is the pulsar spin period in seconds. Since the efficiency 
$f_{\rm prim}$ depends on the NS radius and moment of inertia, we simply 
normalize it by ${\bar f}_{\rm prim}$ to make more illustrative the comparison 
with our calculations performed for a canonical NS model. Note that in 
formula (\ref{fprim2}) the parameter $f_{\rm prim}$ is assumed to be independent 
of $B$ and $P$, but scales with $\kappa $ and $R_6$ in the same way the $\gamma _{\rm acc}$ (or 
the corresponding acceleration potential drop, see eq. [49] in Paper I) does. 

A major advantage of the above formula for $\gamma _{acc}$ is that it does 
not discriminate between unsaturated and saturated regimes of acceleration 
of primary electrons (see HM02). It implies that the Lorentz factor of an accelerating 
electron is merely proportional to the maximum voltage drop above PC, with
the coefficient of proportionality being the bulk efficiency of the pulsar 
accelerator, $f_{\rm prim}$.  

Now we can write the pair-formation condition as 
\be
\gamma _{\rm acc} \geq \gamma _{\rm min},
\label{DLcondition}
\ee 
where $\gamma _{\rm acc}$ is given by formula (\ref{gamma_acc}). 

The expressions for $\gamma _{\rm min}$ for different underlying 
mechanisms for pair-producing photons are the same as derived 
by HM02 but including the explicit dependence on $R$ and $\kappa $, and read 

\noindent{\it Curvature radiation}

%%%%%%%%%%%%%%%%%%%%%%%%%%%%%% CURVATURE %%%%%%%%%%%%%%%%%%%%%%%%%%%%%%%%
\be 
\gamma _{\rm min}^{(CR)} = 10 ^7~   
\left\{ \begin{array}{ll}
    3.4~R_6^{5/14}\kappa _{0.15}^{1/7}P^{1/14}B_{12}^{-1/7} & ~P\lsim P_{\ast }^{(CR)}, \\
    1.2~R_6^{3/4}\kappa _{0.15}^{1/4}P^{-1/4} & ~P\gsim P_{\ast }^{(CR)},
\end{array} 
\right.
\label{gamma_min_CR}
\ee

\noindent{\it Resonant ICS}

%%%%%%%%%%%%%%%%%%%%%%%%%%%%%% RESONANT %%%%%%%%%%%%%%%%%%%%%%%%%%%%%%%%
\be 
\gamma _{\rm min}^{(R)} = 10 ^6~   
\left\{ \begin{array}{ll}
    0.9~R_6^{1/2}\kappa _{0.15}^{1/3}P^{-1/6}B_{12}^{-1} & ~P\lsim P_{\ast }^{(R)}, \\
    0.2~R_6^{5/4}\kappa _{0.15}^{1/2}P^{-3/4}B_{12}^{-1/2} & ~P\gsim P_{\ast }^{(R)},
\end{array} 
\right.
\label{gamma_min_R}
\ee

\noindent{\it Non-resonant ICS}

%%%%%%%%%%%%%%%%%%%%%%%%%%%% NON-RESONANT %%%%%%%%%%%%%%%%%%%%%%%%%%%%%%
\be 
\gamma _{\rm min}^{(NR)} = 10 ^5~   
\left\{ \begin{array}{ll}
    R_6^{1/2}\kappa _{0.15}^{1/3}P^{-1/6}B_{12}^{-1/3} & ~P\lsim P_{\ast }^{(NR)}, \\
    0.6~R_6^{5/4}\kappa _{0.15}^{1/2}P^{-3/4} & ~P\gsim P_{\ast }^{(NR)},
\end{array} 
\right.
\label{gamma_min_NR}
\ee

where
\be
P_{\ast }^{(CR)} = 0.1~B_{12}^{4/9}, 
\ee 
\be
P_{\ast }^{(R)} = 0.1~B_{12}^{6/7},
\ee 
\be
P_{\ast }^{(NR)} = 0.4~B_{12}^{4/7}
\ee
are the critical spin periods (see eqs [4]-[6] in HM02) in 
the criterion defining the unsaturated (upper row) and 
saturated (lower row) regimes of primary electron acceleration. 

Now, let us use formula (\ref{gamma_acc}) in criterion 
(\ref{DLcondition}) to get explicit conditions representing the death 
lines. Note that, following the reasoning of HM02, in criterion 
(\ref{DLcondition}) we should evaluate $\gamma_{acc}$ at 
${\bar f}_{\rm prim}={\bar f}_{\rm prim}^{\rm min}$, where 
${\bar f}_{\rm prim}^{\rm min}$ is the minimum pulsar efficiency 
needed for pair formation. It is this minimum or threshold value of 
$f_{\rm prim}$ that determines the pulsar death line condition. 
We must note that at the pulsar death line, in the case of ICS, the 
$f_{\rm prim}^{\rm min}$ defines the voltage drop at the PFF not the 
final energy of the primary beam (see Section 3). For 
each of the mechanisms of generation of pair-producing photons, we discuss 
in this paper, the 
resultant analytic death line (or rather parameter space with 
allowed pair formation) in the $P$--$\dot P$ diagram reads 

\noindent{\it Curvature radiation}

%%%%%%%%%%%%%%%%%%%%%%%%%%%%%% CURVATURE %%%%%%%%%%%%%%%%%%%%%%%%%%%%%%%%
\be 
\lg {\dot P} \geq ~   
\left\{ \begin{array}{ll}
    {21\over 8}~\lg P -{7\over 4}\lg {\bar f}_{\rm prim}^{\rm min} - 
    \Delta _{\rm I}^{(CR)}(R,I) -14.6 & ~P\lsim P_{\ast }^{(CR)}, \\
    {5\over 2}~\lg P - 2\lg {\bar f}_{\rm prim}^{\rm min} - 
    \Delta _{\rm II}^{(CR)}(R,I)  -15.4 & ~P\gsim P_{\ast }^{(CR)},
\end{array} 
\right.
\label{DL_CR}
\ee
where $\Delta _{\rm I}^{(CR)}(R,I) = 
1.5 (\lg I_{45}-6.3 \lg R_6)$, 
and $\Delta _{\rm II}^{(CR)}(R,I) = 
1.5 (\lg I_{45}-7.3 \lg R_6)$.

\noindent{\it Resonant ICS}

%%%%%%%%%%%%%%%%%%%%%%%%%%%% RESONANT %%%%%%%%%%%%%%%%%%%%%%%%%%%%%%
\be 
\lg {\dot P} \geq ~   
\left\{ \begin{array}{ll}
    {5\over 6}~\lg P - \lg {\bar f}_{\rm prim}^{\rm min}- {2\over 3} 
	\Delta _{\rm I}^{(ICS)}(R,I) -16.6 & ~P\lsim P_{\ast }^{(R)}, \\
    {2\over 3}~\lg P -{4\over 3}\lg {\bar f}_{\rm prim}^{\rm min}- {2\over 3}
	\Delta _{\rm II}^{(ICS)}(R,I)-17.9 & ~P\gsim P_{\ast }^{(R)},
\end{array} 
\right.
\label{DL_R}
\ee

\noindent{\it Non-resonant ICS}

%%%%%%%%%%%%%%%%%%%%%%%%%%%% NON-RESONANT %%%%%%%%%%%%%%%%%%%%%%%%%%%%%%
\be 
\lg {\dot P} \geq ~   
\left\{ \begin{array}{ll}
    {7\over 4}~\lg P -{3\over 2}\lg {\bar f}_{\rm prim}^{\rm min}-
	\Delta _{\rm I}^{(ICS)}(R,I)-18.6 & ~P\lsim P_{\ast }^{(NR)}, \\
    {3\over 2}~\lg P -2\lg {\bar f}_{\rm prim}^{\rm min}- 
	\Delta _{\rm II}^{(ICS)}(R,I)-20.0 & ~P\gsim P_{\ast }^{(NR)},
\end{array} 
\right.
\label{DL_NR}
\ee
where $\Delta _{\rm I}^{(ICS)} = (\lg I_{45}-7.5\lg R_6)$ and  
$\Delta _{\rm II}^{(ICS)} = (\lg I_{45}-10.5\lg R_6)$. In 
the above death-line conditions we used formula (\ref{kappa}) for $\kappa $.

The expressions (\ref{DL_CR})-(\ref{DL_NR}) differ from the similar expressions 
presented in HM02 (see eqs [52]-[54]) by the terms $\Delta $'s which take
into account the deviation of NS radius and moment of inertia from the canonical 
values of $10^6$ cm and $10^{45}$ g$\cdot$cm$^2$, respectively. Thus, for canonical 
NS parameters, the above expressions translate into expressions [52]-[54] of HM02. 
One can see from expressions for $\Delta$'s that the more compact the NS is, the lower 
the death line moves. 

The analytic expressions above for the ICS pair death lines differ significantly from
those derived by Zhang et al. (2000).  The reasons for these differences were discussed
in detail in HM02.

\subsection{Numerical death lines}

\subsubsection{Effect of NS equation of state}

The details of the method we use to numerically compute pair death lines can be found 
in HM02.  Briefly, we keep track of the total distance, the sum of the acceleration
length and the pair production attenuation length of either CR or ICS radiated photons,
as the primary electron is accelerating.  The minimum of this total distance is assumed
to determine the height of the pair formation front.  As the value of surface magnetic 
field decreases for a given pulsar period, the PFF moves to higher altitude. 
Performing this calculation for a range of pulsar periods, we find the value of surface 
magnetic field below which a PFF cannot form within the pulsar magnetosphere.  This occurs
because both the required acceleration length and the pair attenuation length become 
very large.  The result is a line in $P$-$B_0$ space which we identify as the death
line for pair production by photons of a given radiation type.  In order to compare
death lines for different equations of state (EOS) with the observed pulsar population we
must transform the calculated lines to $P$-$\dot P$ space using the magneto-dipole 
spin-down relation 
\be
B_0 = \left[{3c^3I\dot P P\over 2\pi^2 R^6}\right]^{1/2},
\ee
giving
\be
\dot P = 2.43 \times 10^{-16} \,\left({R_6^6\over I_{45}}\right)\,
\left({B_{12}^2\over P}\right) ~~~{\rm s}\cdot {\rm s}^{-1} .
\ee

In our numerical death line calculations we employ three most representative 
NS models standardly used in the calculations of thermal evolution of 
NSs (see e.g. Table 1 in Umeda et. al 1993 and references quoted therein) plus 
a strange star model (see e.g. {Glendenning, 1997). The NS  
models correspond to a star with the baryon mass 1.4 $M_ {\sun }$, and different 
radii and moments of inertia: $R_6 = 1.6$ and $I_{45} = 2.2$ 
(Pandharipande-Pines-Smith'76 model); $R_6 = 1.1$ and $I_{45} = 1.2$ 
(Friedman-Pandharipande'81 model without pion condensate); and $R_6 = 0.8$ and 
$I_{45} = 0.6$ (Baym-Pethick-Sutherland'71 model). The strange star model has 
a mass 1.4 $M_{\sun }$, radius $R_6 = 0.7$, and moment of inertia $I_{45} = 0.7$. 
In Figure 1 we show the death line calculations based on these models. Note that the 
stellar models were produced for a non-rotating star. Thus, strictly speaking, our 
calculations of death lines shown in Figure 1 are not very accurate for the ms pulsars. 
However, as it will be discussed in the next Section our death line calculations based 
on a non-rotating NS (and perhaps strange star) model may still be satisfactory even in 
the ms range. We should also mention that the only purpose of our inclusion of a 
strange star model is to demonstrate the effect of extreme stellar compactness on our  
death line calculations. The surface physics of a strange star may be significantly 
different from that of a NS, and in this paper we refrain from any speculation on 
this issue. It was suggested although that some radio pulsars could well be strange stars 
rather than NSs (see e.g. Xu, Qiao \& Zhang 1999; Kapoor \& Shukre 2001). 

As was suggested by our analytic expressions in Section 2.2, EOS with smaller radii
will move the death line lower, allowing a greater number of pulsars to produce pairs.
Pair production is thus facilitated in more compact stars with bigger $\kappa $'s and 
having softer EOS or even having strange matter EOS. We demonstrated this effect by 
employing the BPS NS model and more or less typical strange star model available 
in the literature. The death line corresponding to the strange star model (having 
largest compactness parameter) is the lowest 
one of those shown in Figure 1. Note, that the ICS pair death lines are more strongly 
affected by a change in NS radius than are the CR pair death lines.  In this paper we
have computed ICS death lines for only one PC temperature of $10^6$ K in order
to compare the effect of EOS.  In HM02, we showed that PC temperature has only a small 
effect on ICS pair death lines for normal pulsars and is much less 
significant than the effect of different NS EOSs.  For a PC temperature of $5 \times 10^6$ K
and canonical NS model, the ICS pair death line lies slightly below the BPS model death line
shown in Figure 1.   

\subsubsection{Effect of NS mass for death lines in ms PSRs}

In our calculations of death lines for pulsars with the periods in the 
range of 0.001-0.1 s we use rapidly rotating NS models produced by Friedman, 
Ipser \& Parker (1986). In Figure 2 we present our calculated death lines 
for NSs with the gravitational masses 1.26, 1.97, and 2.64 $M_{\sun }$. This  
particular sequence of rotating NS models is calculated by employing the  
Pandharipande-Smith'75 EOS and corresponds to the NS spin period of 
$\approx 2$ ms. The NS radii and moments of inertia for this sequence are, 
respectively, $R_6 = 1.59$ and $I_{45} = 2.28$, $R_6 = 1.59$ and $I_{45} = 3.9$, 
and $R_6 = 1.47$ and $I_{45} = 5.18$. Note that we used the same sequence of models 
(implying the NS spin period $\approx 2$ ms) to calculate the death lines for 
the whole range of spin periods up to 0.1 s. In fact, for the spin periods 
$\gsim 10$ ms the rotating NS models practically converge with the non-rotating 
models. Note also that, for these three models the relative differences between 
non-rotating and rotating sequences in terms of, respectively, the gravitational 
mass, radius, and moment of inertia, are 3$\% $, 10$\% $, and 8$\% $ 
(for 1.26 $M_{\sun }$ model); 1.5$\% $, 5$\% $, and 5$\% $ (for 1.97 $M_{\sun }$ model); 
and 6$\% $, 8$\% $, and 4$\% $ (for 2.64 $M_{\sun }$ model). The differences between 
rotating and non-rotating models of this magnitude are more or less typical for 
other sequences of models based on a reasonable EOS. Thus, the effect of rotation, 
by itself, is not very important for the death line calculations, and we can justifiably 
use this particular sequence of models for our death line calculations for the 
period range under consideration. What may actually be important for the death line 
calculations in ms pulsars is the mass of a NS. We find that our 
numerical calculations of death lines in ms pulsars favour more massive NS 
models, which is consistent with the hypothesis that the ms pulsars descend 
from accreting NSs in low-mass binaries. Figure 2 illustrates the effect of NS 
mass on the death line calculations for the ms pulsars. It shows that the 
increase in NS (gravitational) mass by 0.6-0.7 $M_{\sun }$ moves the death line 
down by a factor of a few or more (see also formula [\ref{DL_NR}], saturated case).  
Even though the mass of 2.64 $M_{\sun }$ we use in our calculations may seem to be rather 
large, the main result of our calculations shown in Figure 2 is that the increase in 
the NS baryon mass by $\sim 30-60\% $ may significantly 
facilitate the process of pair formation above the PC in a ms pulsar. This effect 
may account for the fact that many ms pulsars tend to scatter below the theoretical death lines. 
The fact that the ms pulsars might be more massive NSs processed in binary systems could 
naturally explain this effect.

\subsubsection{Effect of two-photon pair formation}

We also investigate the effect of two-photon pair production on the pair death line.
The process we consider is that of ICS photons interacting with soft X-ray 
photons from the hot PC, drawing from the same pool of thermal photons that are 
responsible for creating the ICS photon spectrum.  In PC models, two-photon pair 
production has traditionally not been considered important in comparison to one-photon pair
production.  Zhang \& Qiao (1998) noted that two-photon pair production could be
important in PC models if the temperature of the PC was high enough ($> 4 \times 10^6$ K 
for $P = 0.1$ s).   Zhang (2001) estimated the photon attenuation length for two-photon 
pair production above a hot PC of radius $R_t$ to be
\be
\ell_{2\gamma} \simeq 4.7 \times 10^5\,T_6^{-3}\,[g(z)\Sigma (\epsilon)]^{-1}, {\rm cm}
\ee
where 
\begin{eqnarray} \label{muc}
g(z) & = & 0.27 - 0.507\mu_c + 0.237\mu_c^2, \\ \nonumber
\mu_c & = & {z\over \sqrt{z^2 + z_t^2}},
\end{eqnarray}
where $z_t = R_t/R$, $T_6 = T/10^6$ K, 
$\Sigma (\epsilon) = (\pi^2/3)\ln(0.117\Theta\epsilon)/(\Theta\epsilon)$,  
$\Theta = T/mc^2$, and $\epsilon$ is the photon energy in units of $mc^2$.  
Near the NS surface $\mu_c \sim 0$, and at threshold,
\be \label{eps_th}
\epsilon_{th} = {2\over\Theta (1-\mu_c)} ,
\ee
where  $\epsilon \sim 1/\Theta$, $\Sigma (\epsilon) \sim 1$ so that 
$\ell_{2\gamma} \simeq 1.7 \times 10^6\,T_6^{-3}\,{\rm cm}$.  For surface temperatures
$T_6 \sim 1$, the photon attenuation length, $\ell_{2\gamma}$, is much larger than the
acceleration length required for the electron to radiate an ICS photon above threshold.
Therefore, $\ell_{2\gamma}$ sets the distance to the two-photon PFF.  Since the soft photon
density declines with height above the surface on a scale roughly equal to $R_t$,
a reasonable criterion for two-photon pair production is then $\ell_{2\gamma} < R_t$.
Since $R_t = r_{pc} = (\Omega R/c)^{1/2}R$ for a heated PC, this condition becomes
\be
T_6 \gsim 1.6 (P/1~{\rm ms})^{1/6}\,R_6^{1/2}
\ee 
For normal pulsars, $T_6 > 4-5$ is required, which is unrealistically high, but for ms 
pulsars the temperature required for significant two-photon pair production is in
the range detected for some ms pulsars.  It is clear that the advantage ms pulsars
hold over normal pulsars in the facilitation of two-photon pair production is a
large PC size, which allows both larger angles between the ICS $\gamma$-rays and 
the thermal PC photons thus lowering the threshold energy for producing a pair,
and an increase in the scale length over which the soft photon density decays.  
The primary electrons therefore can reach the energies needed to radiate photons
at threshold in a shorter distance.  

In order to compute the two-photon pair death lines numerically, we need an 
expression for the rate of pair production of a high energy photon of energy
$\epsilon$:
\be \label{R2g1}
R_{2\gamma}(\epsilon, \theta) = c\int d\phi \int d\mu_{ti} \int d\epsilon_s
\sigma_{2\gamma}(w)\,n_s(\epsilon_s, \mu_s)\,(1-\mu_{ti}),
\ee
where $\mu_{ti}$ is the cosine of the polar angle between the propagation direction 
of the two photons, and $\sigma_{2\gamma}(w)$ is the cross section,
\be
\sigma_{2\gamma}(w) = {\pi r_e^2\over w^6}\,\left[(2w ^4 + 2w^2 - 1)
\ln(w + \sqrt{w^2-1})- w (w^2+1)\sqrt{w^2-1}\right]
\ee
in the center of momentum frame in terms of the variable,\
\be
w = [\epsilon\,\epsilon_s(1-\mu_{ti})/2]^{1/2}.
\ee
The above cross section does not take into account the effect of the strong magnetic 
field near the NS surface.  Although these effects may be significant in the
highest pulsar fields, the magnetic two-photon pair production cross section is 
very complicated (Kozlenkov \& Mitrofanov 1986) and, since the process will only
be important for ms pulsars having low fields, we will not consider these effects
here.
The above (field-free) cross section may be simplified in two limits, near threshold 
and for large $w$ (Svensson 1982):
\be \label{sig}
\sigma_{2\gamma}(w) \simeq 
\left\{ \begin{array}{ll}
\pi r_e^2\,\sqrt{w ^2 - 1} & w \simeq 1 \\
(\pi r_e^2/w^6)\,[2\ln(2w) - 1] & w \gg 1 ,
\end{array}
\right.
\ee
where $r_e$ is the classical electron radius.
We choose the coordinate system so that the z-axis is along the magnetic pole.  
To simplify the geometry of the calculation, we assume that the $\gamma$-ray travels
along the positive z-axis, and assume that the soft photons are uniformly radiated
over the PC.  There is thus azimuthal symmetry about the magnetic
pole and the polar angle $\mu_{ti}$ ranges from $0$ to $\mu_c$, where $\mu_c$
is given in equation (\ref{muc}).  The thermal photons from the PC are described by the
blackbody distribution,
\be
n_s(\epsilon_s) = (1-\mu_c){8\pi\over \lambda_c^3}\,{\epsilon_s^2\over 
[\exp(\epsilon_s/\Theta)-1]},
\ee
where $\lambda _c$ is the electron Compton wavelength. Changing variables from $\mu_{ti}$
to $w$,
and using the expressions for $\sigma_{2\gamma}(w)$ defined by equation (\ref{sig}), the
expression for the rate in equation (\ref{R2g1}) becomes
\be \label{R2g}
R_{2\gamma}(\epsilon) \simeq 16\pi^2 r_e^2\,{c\over \epsilon^2}\,\int_0^{\infty}d\epsilon_s
\,{n_s(\epsilon_s)\over \epsilon_s^2}\,\Sigma (w_s)
\ee
where
\be
\Sigma (w_s) =
\left\{ \begin{array}{ll}
[{1\over 5}(w_s^2 - 1)^{5/2} - {1\over 3}(w_s^2 - 1)^{3/2}] & w_s \simeq 1 \\
& \\
\left[2.39 - \mbox{\large{$
{2\ln{2w_s}\over w_s} - {1\over w_s}$}} \right] & w_s \gg 1
\end{array}
\right.
\ee
and
\be
w_s = \max[1, \,\epsilon\,\epsilon_s\,(1-\mu_c)/2]. 
\ee 

Equation (\ref{R2g}) is then integrated numerically to obtain the two-photon pair production
and attenuation length.

As in the case of the one-photon PFF calculation, we minimize the 
sum of the acceleration length and the pair production attenuation length of ICS radiated 
photons, as the primary electron is accelerating.  Performing this calculation for a 
range of pulsar periods, we find the value of surface magnetic field below which a PFF 
cannot form within the pulsar magnetosphere.  Figure 3 shows the computed pair death lines
in $P$-$\dot P$ space that include the possibility of two-photon pair production for 
different values of the PC surface temperature.  We display three cases for illustration:
1) death lines for one-photon pairs only 2) death lines for two-photon pairs only and 
3) death lines for one-photon and two-photon pairs.  All cases assume a canonical NS model
with $M = 1.4 M_{\sun}$, $I_{45} = 1$ and $R = 10$ km.  It is apparent that two-photon
pair production is not important at all for any of the known radio pulsar population
unless the PC temperature $T_6 \gsim 3$.  The position of the two-photon death line is 
sensitively dependent on PC temperature for $3.0 \lsim T_6 \lsim 5.0$ and then saturates
at about $T_6 \sim 5.0$, reflecting the effect of the two-photon pair threshold.  For
$T_6 \lsim 3.0$, the ICS photons never reach pair threshold during the particle acceleration.
For $3.0 \lsim T_6 \lsim 5.0$, the photons are pair producing near threshold where the
cross section is sharply rising, and for $T_6 \gsim 5.0$ the photons are pair producing 
above threshold where the cross section is decreasing.  The two-photon death lines curve 
upward to become almost vertical with increasing $P$ because for longer periods, particles
must accelerate to high altitudes to reach pair threshold where the thermal photon density
is declining.  Thus, as we had noted previously, two-photon pair production is only
important in short-period and ms pulsars.  Since young, short-period pulsars with high
magnetic fields do not have detected PC surface temperatures as high as $T_6 \sim 3$, 
two-photon pairs are effectively not important for any but ms pulsars.  The combined one-photon
plus two-photon death lines blend into the one-photon death lines as one-photon pairs
dominate at higher fields and longer periods.

HM02 found that substantial PC heating by trapped positrons returning from an ICS
pair front can occur in ms pulsars if PC temperatures exceed $T_6 \sim 1$.  In order
for ms pulsars to sustain these high temperatures through PC heating, the heated area
must be much smaller than the area of the standard PC, which is $A_{pc} = \pi R^2
(\Omega R/c)$.  This is in fact consistent with the non-uniform heating distribution
found by HM02.  However, for the PC temperatures $T_6 > 3$ needed for two-photon
pair production the question of the stability of two-photon PFFs must be addressed.
Positrons returning from the PFF will radiate ICS photons which can produce two-photon 
pairs in a relatively small distance because the pair production
threshold (see eq. [\ref{eps_th}]) is much lower for head-on collisions.  Creation of
enough two-photon pairs by the returning positrons at high altitudes could disrupt
the acceleration of the primary electrons.  Investigation of this effect will require
inclusion of full angular dependence of the pair production rate and will be 
considered in a future paper.

\section{Acceleration and $\gamma-$ray luminosity}

Establishing the regimes of pair formation above pulsar PCs is not only important
for understanding the behavior of the radio emission, but also allows us to 
predict regimes of particle acceleration and thus high-energy emission since the
acceleration of the primary particles may be limited by screening at a PFF.  
HM02 found that CR pairs are very effective in screening the electric field at the
PFF, whereas ICS pairs are less effective and may only screen the electric field
above the PFF in some cases.  They found that when ICS screening does occur, it only
screens the field locally but will not screen at higher altitudes.  Thus ICS pairs
may retard but do not ultimately limit acceleration of the primary electrons, which
may then also produce CR pairs at higher altitude.  In fact, the luminosity of the 
thermal X-rays from a hot PC detected from PSR B1929+10, a 3 Myr old pulsar where a detectable cooling component is not expected, would require and is consistent with heating by positrons produced at a CR pair front (HM01) since 
the heating by positrons produced only at a ICS pair front would not be detectable 
(HM02).   

The luminosity of the primary electron beam in the PC pulsar model 
can be calculated as
\be
L_{\rm prim} = \alpha c \int | \rho _e | \Phi d S ,
\label{Lprim}
\ee
where 
\be
|\rho _e| = {{\Omega B_0 }\over {2\pi c \alpha \eta ^3}}
{f(\eta)\over f(1)}(1-\kappa) ,
\label{rhoe}
\ee 
is the value of an electron charge density calculated at 
$\cos \chi \approx 1$ (where $\chi $ is the pulsar obliquity), 
$\Phi (z,\xi ,\phi )$ is the electric potential, and 
\be
d S = {{\Omega R ^3}\over {cf(\eta )}} \eta ^3 \xi d \xi d \phi 
\label{dS}
\ee
is the element of a spherical surface cut by the last open field lines 
at the radial distance r ($=R\eta $). Here  $\alpha = \sqrt{1-r_g/R}$ , 
$r_g$ is the gravitational radius of a NS; $c$ is the velocity of light; 
 $z$ is the altitude above the PC 
in units of stellar radius; $\xi $ is the magnetic colatitude of a field 
line scaled by the  magnetic colatitude of the last open field line; and 
$\phi $ is the magnetic azimuthal angle.

In our previous papers (see e.g. HM02) we calculated $L_{\rm prim}$ 
using in formula (\ref{Lprim}) the expression for the electric potential 
evaluated at the relatively smaller altitudes (both for the unsaturated and 
saturated regimes of acceleration of primaries) where the bulk of the 
CR pair-formation and electric field screening occur. In the regime where CR pairs are created,
i.e. above the CR death line, the luminosity of the primary beam is therefore set by the 
CR pair front.  HM02 derived the following expressions for the luminosity 
of the primary electron beam based on the altitude of the CR pair front,
\be \label{LprimCR}
L_{\rm prim}^{(CR)} = 10^{16}~{(\rm erg/s)^{1/2}}~~{\dot E}_{\rm rot}^{1/2}~~ 
\left\{ \begin{array}{ll}
    P^{1/14}B_{12}^{-1/7} &  P\lsim P_{\ast }^{(CR)}, \\
    0.3~P^{-1/4} & P\gsim P_{\ast }^{(CR)}.
\end{array} 
\right.
\ee
In the case where there are no pairs produced by CR (and therefore no electric 
field screening) the most appropriate expression for the electric potential in this case 
is  (see eq. [24] in HM01, and eq. [13] in HM98)
\be
\Phi _{|_{\cos \chi \approx 1}}= 
\Phi _0 {{\Omega R}\over {f(1)c}} (1-\xi ^2)\kappa  
\left\{ \begin{array}{ll}
    {3\over 2}z &  r_{pc}/R \ll z < 1, \\
    {1\over 2}\left( 1- {1\over \eta ^3}\right) & \eta -1 \geq 1 ,
\end{array} 
\right.
\label{Phi}
\ee
where $\eta = r/R$. This formula applies for the altitudes much greater than the PC 
radius and corresponds to the saturated and unscreened regime of acceleration of 
primaries. However, in some cases the acceleration of primary electrons may be 
limited by CR losses, where general formula (\ref{Lprim}), that does not take into 
account the radiation reaction of accelerating particles, may not be 
applicable. 

After substituting expression (\ref{rhoe}) for $|\rho _e|$ and the above 
expression for $\Phi $ into formula (\ref{Lprim}), and performing integration 
over $\xi $ and $\phi $, we get
\be
L_{\rm prim} (\eta )_{|_{\cos \chi \approx 1}} = 
{3\over 4} \kappa (1-\kappa ) 
\left( 1- {1\over \eta ^3} \right) {\dot E}_{\rm rot} .
\label{Lprim,2}
\ee
Formally, in the above formula we should put $\eta \rightarrow \infty $ to calculate 
the maximum power in the primary beam. Thus, for the maximum power of the primary beam 
we can write (see MH97, eq. [76])
\be
L_{\rm prim, max} = {3\over 4} \kappa (1-\kappa ) {\dot E}_{\rm rot} .
\label{Lprim,3}
\ee
To estimate the pulsar bolometric photon luminosity it is reasonable  
to assume that $L_{\gamma} \approx 0.5 L_{\rm prim, max}$, where 
$L_{\rm prim, max}$ is given by formula (\ref{Lprim,3}). Then, using 
expression (\ref{kappa}) for $\kappa $ we can write  
\be 
L_{\gamma } \approx 0.05 {{I_{45}}\over {R_6^3}}
\left( 1-0.15{{I_{45}}\over {R_6^3}}\right) {\dot E}_{\rm rot} .
\label{Lgamma}
\ee
Thus, the energetics of CR photons generated by accelerating 
electrons above the pulsar PC is proportional to the pulsar 
spin-down luminosity, and, according to formula (\ref{Lgamma}),  
the maximum efficiency of conversion of pulsar spin-down power into 
the high-energy quanta may amount to 10$\% $. Note that this 
estimate of efficiency implies that only half of the power 
of accelerating electrons gets consumed by $\gamma $-ray photons.

In Figure 4 we plot the predicted high-energy luminosity as a function of spin-down
luminosity.  In the CR pair regime, where eq. (\ref{LprimCR}) applies, 
we have plotted the luminosity
for the screened unsaturated case (see top eq.[\ref{LprimCR}]).  Below the CR death line, 
${\dot E}_{rot} \lsim 10^{34}\,{\rm erg\,s^{-1}}$,
the luminosity for the unscreened case of equation (\ref{LprimCR}) applies.  Also plotted are 
the luminosities of the pulsars with detected high-energy emission and their predicted luminosities. 
The detected high-energy pulsars are all above the CR death line, although Geminga and
PSR B1055-52 are just above the line.  The detected high energy pulsars seem to follow the
predicted relationship, $L_{\gamma} \propto {\dot E}_{rot}^{1/2}$.  We predict that this relationship
will break to $L_{\gamma} \propto {\dot E}_{rot}$ at $E_{rot} \lsim 10^{34}\,{\rm erg\,s^{-1}}$,
so that older pulsars will have lower predicted luminosities than what would be predicted
by an extrapolation of the trend seen in the younger pulsars.  Non-thermal high-energy
emission has in fact not been detected from older nearby pulsars such as PSR B1929+10
and PSR B0950+08, although a thermal emission component has been detected from PSR B1929+10
(Wang \& Halpern 1997) which may be due to PC heating (see HM01).  
EGRET upper limits for pulsed emission from these pulsars (Thompson et al.
1994) are plotted in Figure 4 and lie above the predicted luminosities, but not by much.
GLAST should be able to detect $\gamma $-ray emission from these older pulsars and test 
the predicted $L_{\gamma} \propto {\dot E}_{rot}$ dependence and its location. Note that 
the break value of ${\dot E}_{\rm rot}$ in the $L_{\gamma }({\dot E}_{\rm rot})$ dependence 
depicted in Figure 4, can be estimated by equating the screened expression for 
$L_{\gamma} \approx 0.5 L_{\rm prim }^{(CR)}({\dot E}_{\rm rot})$ given by formula (\ref{LprimCR}) 
at $L_{\gamma ,{\rm max}} \approx 0.5 L_{\rm prim, max}$, where $L_{\rm prim, max}$ 
is given by formula (\ref{Lprim,3}),
\be
5\times 10^{15} {\dot E}_{\rm rot}^{1/2} P^{-1/4} = 0.04~{\dot E}_{\rm rot }, 
\ee
which gives
\be
{\dot E}_{\rm rot, break} = 1.4 \times 10^{34}~P^{-1/7}\,B_{12}^{-2/7}~~~~~{\rm erg}\cdot {\rm s}^{-1}.
\ee
For $P \approx 0.1$ s, $B_{12} \approx 4$, which is shown in Figure 4, 
${\dot E}_{\rm rot, break} \approx 6.6 \times 10^{33}$ erg$\cdot $s$^{-1}$. Because of the $B_{12}^{-2/7}$ dependence 
${\dot E}_{\rm rot, break}$ should generally occur at higher ${\dot E}_{\rm rot}$ values for ms pulsars. 

There are a number of ms pulsars which, according to equation (\ref{LprimCR}), should have
observable high-energy emission.  Several of these, including PSR J0437-4715, 
PSR J0030+0451, PSR J1824-2452 and PSR J0218+4232, 
have non-thermal pulsed X-ray emission components but have not, with the possible
exception of PSR J0218+4232 (see Kuiper et al. 2000), been detected in the 
$\gamma$-ray band. It is important to remember, however, that the predicted luminosity   
$L_{\gamma }$ is a bolometric luminosity.  The luminosity in a particular band is
very dependent on the actual emission spectrum.  The $\nu F_{\nu }$ spectra of the bright 
$\gamma$-ray pulsars all peak in the $\gamma$-ray or hard X-ray bands with typical
high-energy turnovers around several GeV.  The high-energy spectrum of the ms pulsars 
may be quite different.  The bulk of the ms pulsars are near or below the CR death line, 
so that their accelerating electric field is unscreened.  The accelerating primary 
electrons will reach an energy where CR losses are compensated by the acceleration 
energy gain (see also Luo et al. 2000),
\be
\gamma_{_{\sc CRR}} = 1.8 \times 10^7\, B_{12}^{1/4}\,\kappa_{0.15}^{1/4}\,P^{-1/4} .
\ee 
The
CR emission spectrum of these electrons will be quite hard (photon index $-2/3$) and
will not be cutoff by magnetic pair production at an energy of a few GeV, but will
extend to the natural cutoff of the CR spectrum at
\be
\epsilon_{_{\sc CR}} = {3\over 2}{\hbar\over mc^2}{\gamma_{_{\sc CRR}}^3\over \rho _c}
= 2.3 \times 10^5 \left({B_{12}\,\kappa_{0.15}\over P}\right)^{3/4},
\ee
where $\rho _c $ is the radius of curvature of a dipole field line.
The CR $\nu F_{\nu }$ spectrum of the ms pulsars is therefore expected to peak at 50-100
GeV energies.  The numerical model spectra for millisecond pulsars of Bulik et al. (2000) 
have also shown this result.  The high energy curvature radiation from ms pulsars therefore
falls in an energy band that has been above that of satellite detectors like EGRET and below 
that of air Cherenkov detectors.  There could be a second spectral component of almost 
comparable total
luminosity due to the synchrotron pair cascade from ICS pairs, but the $\nu F_{\nu }$
spectrum of this component would peak at much lower energies.  There is also a question of whether two-photon pairs could screen the accelerating field and 
limit the voltage to very low values.  Our preliminary investigations indicate 
that screening by two-photon pairs occurs only for NS surface temperatures high 
enough ($T_6 \gsim 4$) that two-photon pairs from returning positrons is likely to disrupt the acceleration (see discussion at end of Section 2.3.3).  This 
issue will be a subject for future, more detailed investigation.

\section{Summary and conclusions}

In this paper we have outlined the status of two basic regimes of primary 
particle acceleration above a pulsar PC in the general-relativistic version of the 
space-charge limited flow model: the regime of acceleration with subsequent 
electron-positron pair formation by CR and screening of the electric field, and 
the regime of unscreened acceleration with ICS pairs and the 
emission of energetic CR photons. For the pulsar physical parameters we discussed in this 
paper such as the pulsar spin period, NS surface magnetic field strength, 
and PC temperature, the pair production may involve both the one-photon (magnetic) and 
two-photon (mostly $\gamma $-ray ICS photons on thermal X-ray photons) mechanisms, 
with high-energy photons generated via CR and ICS processes. 
In HM02 we began analyzing the onset of pair formation and addressed how the corresponding 
pair-formation criterion transforms into the pulsar theoretical death lines. HM02 presented 
the  death line calculations for the canonical NS model (mass of 1.4 ~$M_{\sun}$, radius 
of 10 km, and moment of inertia $I = 10^{45}~{\rm g}\cdot {\rm cm}^2$) and for one-photon 
pair formation mechanism only. Here we extended our study of pulsar theoretical death lines 
to include the effect of NS mass and radius and, for the ms pulsars with relatively 
hot PCs ($T \sim 3-4 \times 10^6$ K), to incorporate the mechanism of two-photon pair 
formation. In the present study we demonstrate that the effect of NS mass is 
important for the death line calculations in ms pulsars. In fact, if the 
ms pulsars are more massive post-accreting NSs spun-up in low-mass binary 
systems, then the phase space they occupy in the $P-\dot P$ diagram would be consistent 
with our theoretical death lines calculated for NSs with masses more than 1.4~$M_{\sun }$. 

For the regime of acceleration without CR pairs we calculate the 
$\gamma $-ray luminosity of the primary beam as a function of pulsar spin-down 
power. It is important that the form of the theoretical $L_{\gamma }-{\dot E}_{\rm rot }$ 
dependence is determined by the regime of primary acceleration. For example, for the 
regime of acceleration accompanied by pair formation capable of screening the accelerating 
field $L_{\gamma } \propto {\dot E}_{\rm rot }^{1/2}$, whereas for the unscreened regime of 
acceleration $L_{\gamma } \propto {\dot E}_{\rm rot }$. We discuss our theoretical 
$L_{\gamma }-{\dot E}_{\rm rot}$ plots in the light of currently available 
pulsar $\gamma $-ray data, and predict that for pulsars in the regime of 
unscreened acceleration (with relatively low values of ${\dot E}_{\rm rot }$) 
$L_{\gamma }$ should turn down from the ${\dot E}_{\rm rot}^{1/2}$ dependence 
(see Figure 4). 

The main conclusions of our study can be summarized in the following way

\begin{itemize}

\item  We revised the prescription for the derivation of pulsar theoretical death lines 
to include the effect of variation (by more than an order of magnitude) of pair-formation 
altitude with pulsar parameters $P$ and $\dot P$. 

\item  Pulsar theoretical death lines are strongly affected by the EOS of a NS, with the 
onset of pair formation facilitated in more compact NSs. 

\item  Theoretical death lines for ms pulsars produced for massive NS models 
are in a good agreement with the empirical death line and with the hypothesis that 
ms pulsars are descendants of accreting NS in low-mass binaries. 

\item  The ms pulsars with relatively hot PCs (temperature $\sim 3-4 \times 10^6~K$) 
may allow the occurrence of two-photon pair production, which significantly facilitates 
pair formation and may also move down the theoretical death lines in the ms range.

\item  We derive a theoretical expression for $L_{\gamma }$ and illustrate 
that it scales simply as ${\dot E}_{\rm rot }$ or ${\dot E}_{\rm rot }^{1/2}$, respectively, for the 
unscreened regime of acceleration of the primary beam or for the acceleration implying pair formation 
and screening of the accelerating electric field. We predict that the break in $L_{\gamma }- 
{\dot E}_{\rm rot }$ dependence, attributed to the transition from the regime of acceleration 
with pair screening to the unscreened regime, might be seen in observational data.

\end{itemize}

%%%%%%%%%%%%%%%%%%%%%%%%%%%%%%%%%%%%%%%%%%%%%%%%%%%%%%%%%%%%%%%%%%%%%%%%%
\acknowledgments %%%%%%%%%%%%%%%%%%%%%%%%%%%%%%%%%%%%%%%%%%%%%%%%%%%%%%%%
%%%%%%%%%%%%%%%%%%%%%%%%%%%%%%%%%%%%%%%%%%%%%%%%%%%%%%%%%%%%%%%%%%%%%%%%%
We  acknowledge support from the NASA Astrophysics Theory Program.

\newpage
%\centerline{figure Captions}
\figureout{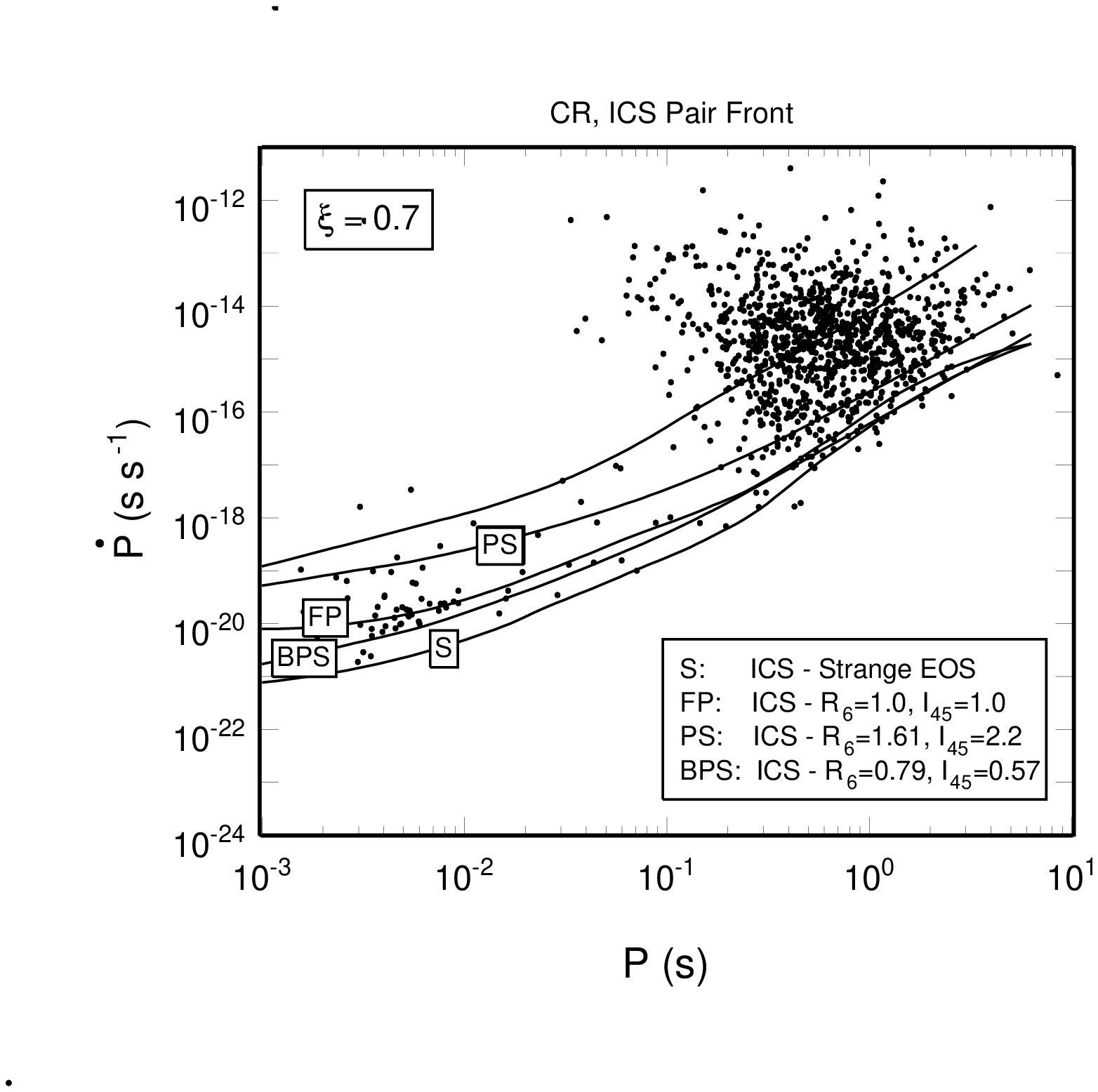}{0}{
Pair death lines in the pulsar $P$-$\dot P$ diagram for curvature radiation (unlabeled
curve) and for inverse-Compton radiation for different NS equations of state (see text). 
The parameter $\xi = \theta/\theta_{pc}$, where $\theta_{pc} = (\Omega R/c)^{1/2}$, 
indicates the magnetic colatitude of the
primary electron acceleration.  Also shown are radio pulsars in the ATNF Pulsar Catalog 
(available at http://www.atnf.csiro.au/pulsar/).
    \label{fig:Fig1} }    %fig. 1
 
\figureout{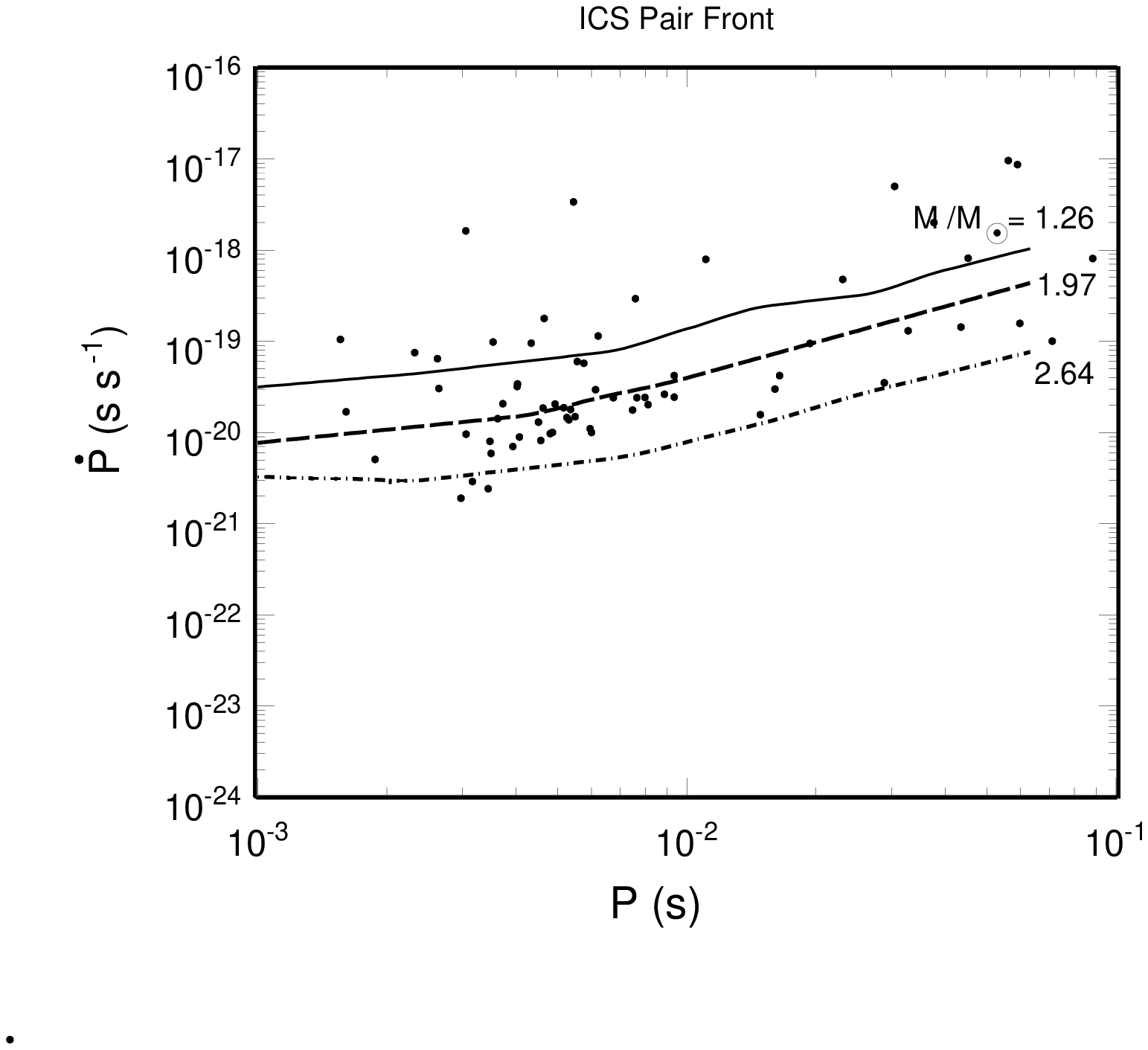}{0}{
Pair death lines for inverse-Compton radiation in the pulsar $P$-$\dot P$ diagram of 
short-period pulsars from the ATNF Pulsar Catalog, for rotating NS models having 
different masses (see text).
    \label{fig:Fig2} }    %fig. 2

\figureout{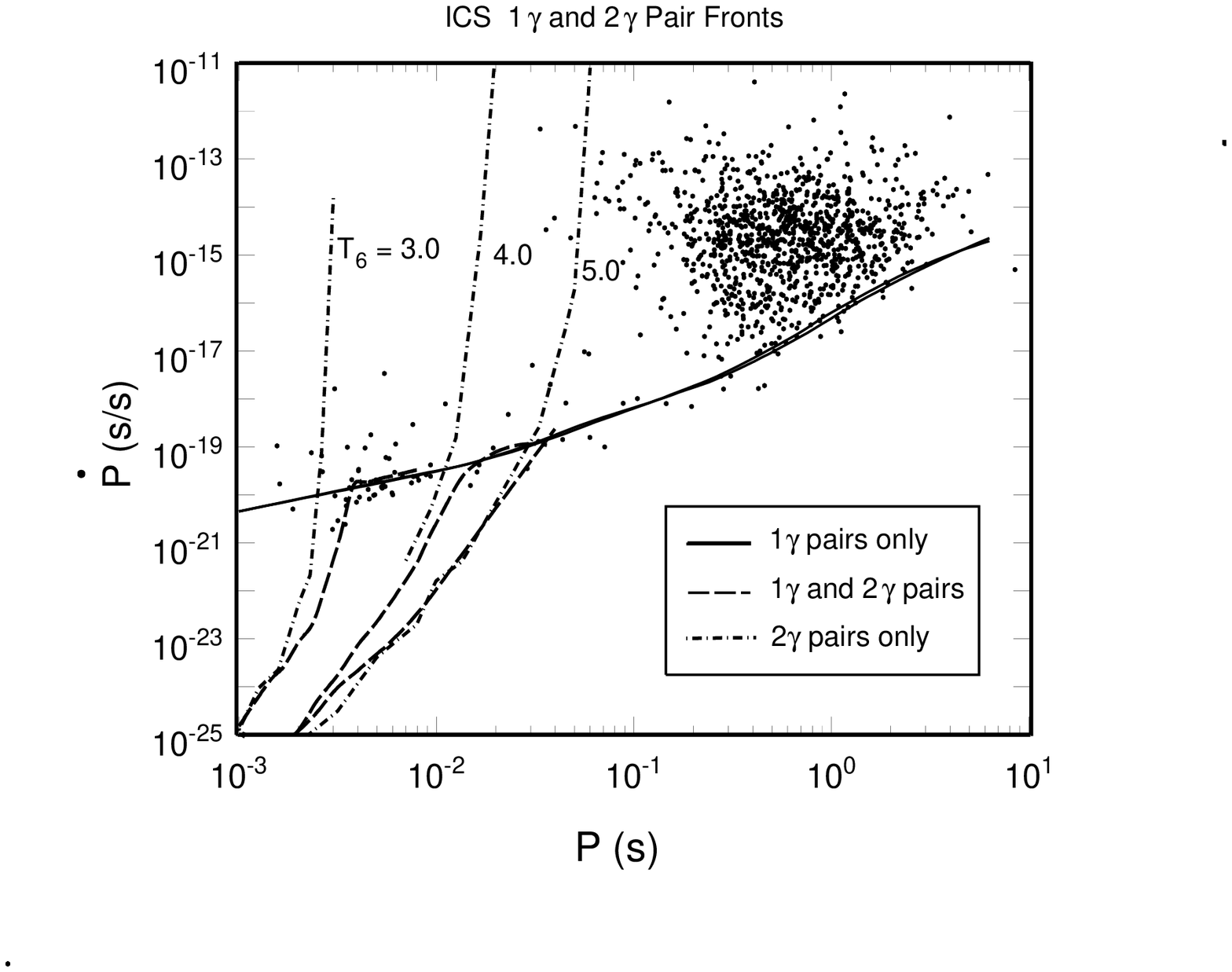}{0}{
One-photon and two-photon pair death lines for inverse-Compton radiation in the pulsar 
$P$-$\dot P$ diagram for different PC surface temperatures, $T_6 = T/10^6$ K, as labeled.  
Also shown are radio pulsars in the ATNF Pulsar Catalog.
   \label{fig:Fig3} }    %fig. 3

\figureout{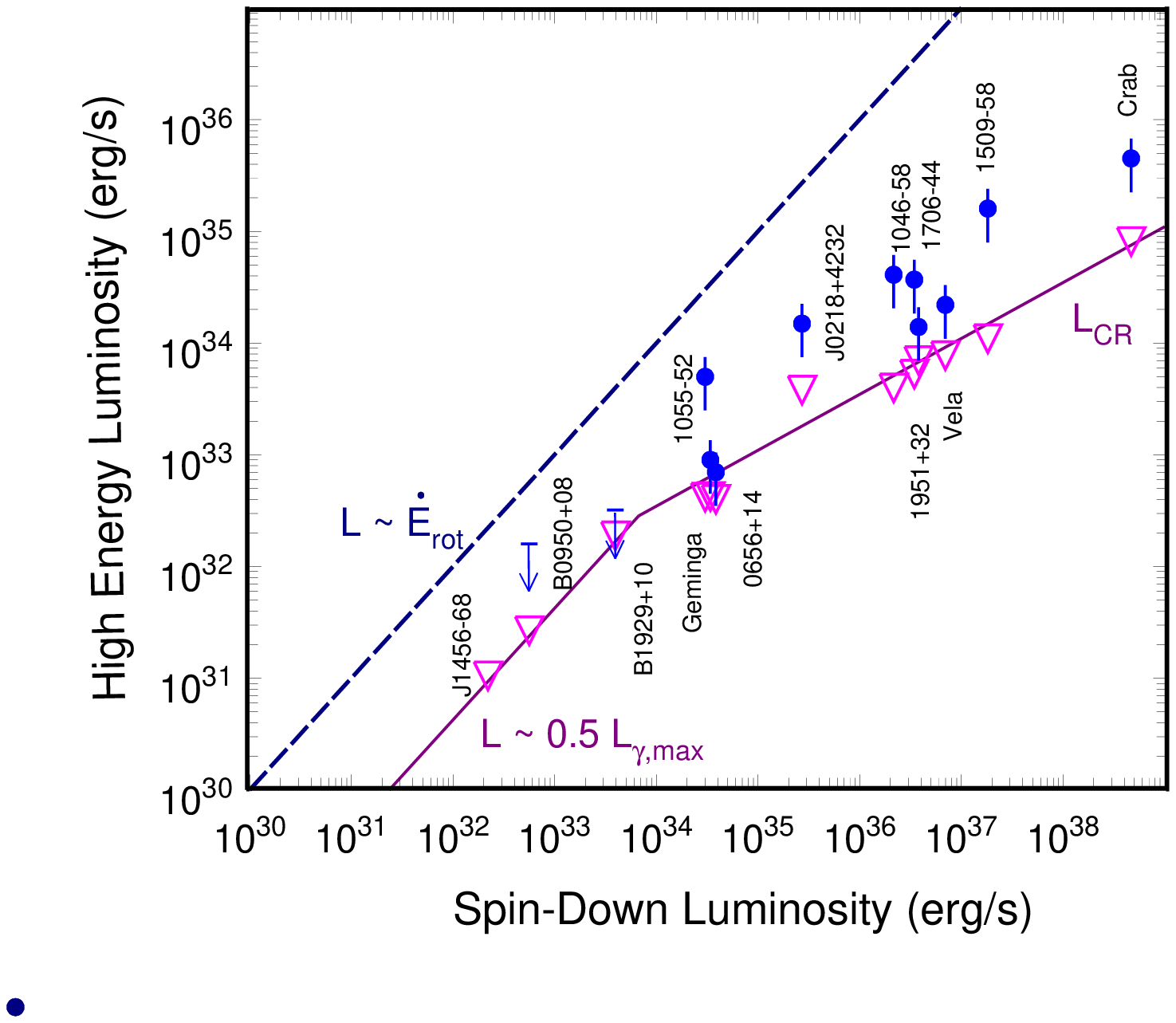}{0}{
Predicted and observed high energy luminosity vs. spin-down luminosity.  The solid
curve is the theoretical prediction from first expression in eq. (\ref{LprimCR}) and 
eq. (\ref{Lgamma}).  
The solid circles are the luminosities of the detected $\gamma$-ray pulsars,  
Thompson (2001), derived from detected fluxes above 1 eV assuming a 1 sr. solid angle. 
The upper limits 
are for $> 100$ MeV from Thompson et al. (1994).  The open triangles are predicted
luminosities for the detected pulsars.
    \label{fig:Fig4} }    %fig. 4
 
%%%UCP%%%

\begin{references}
\reference{}
Arons, J. \& Scharlemann, E. T. 1979, ApJ, 231, 854.
\reference{}
Bulik, T., Dyks, J. \& Rudak, B. 2000, MNRAS, 317, 97.
\reference{}
Chen, K., \& Ruderman, M. 1993, ApJ, 402, 264 (CR93).
\reference{}
Cheng, K. S., Ho, C. \& Ruderman, M. A. 1986, ApJ, 300,500.
\reference{}
Daugherty, J. K. \& Harding, A. K. 1996, ApJ, 458, 278. 
\reference{}
Friedman, J. L., Ipser, J. R., \& Parker, L. 1986, ApJ, 304, 115.
\reference{}
Glendenning, N. 1997, Nuclear and Particle Physics of Compact Stars (Berlin: Springer).
\reference{}     
Harding, A. K., \& Muslimov, A. G. 1998, ApJ, 508, 328 (HM98).
\reference{}
Harding, A. K., \& Muslimov, A. G. 2001, ApJ, 556, 987 (HM01).
\reference{}
Harding, A. K., \& Muslimov, A. G. 2002, ApJ, 568, 000 (HM02).
\reference{}
Hibschman, J. A., \& Arons, J. 2001, ApJ, 554, 624 (HA01)
\reference{}
Kapoor, R. S., \& Shukre, C. S.2001, A\&A, 375, 405
\reference{} 
Kozlenkov, A. A., \& Mitrofanov, I. G. 1986, Sov. Phys.-JETP Lett., 64, 1173.
\reference{}
Kuiper, L., et al. 2000, A \& A, 359, 615.
\reference{}
Luo, Q., Shibata, S. \& Melrose, D. B. 2000, MNRAS, 318, 943.
\reference{}
Manchester, R. N. et al. 2001, MNRAS, 328, 17.
\reference{}
Muslimov, A. G., \& Tsygan, A. I. 1992, MNRAS, 255, 61 (MT92).
\reference{}
Muslimov, A. G., \& Harding, A. K. 1997, ApJ, 485, 735 (MH97).
\reference{}
Romani, R. W., 1996, ApJ, 470, 469.
\reference{}  
Ruderman, M., \& Sutherland, P. G. 1975, ApJ, 196, 51.
\reference{}
Sturrock, P. 1971, ApJ, 164, 529.
\reference{}
Svensson, R. 1982, ApJ, 258, 335.
\reference{} 
Thompson, D. J., et al. 1994, ApJ, 436, 229. 
\reference{} 
Thompson, D. J., 2001, in High-Energy Gamma-Ray Astronomy, ed. F. A. Aharonian \& H. J. 
Volk, (AIP, New York), p. 103.
\reference{}
Umeda, H., Shibazaki, N., Nomoto, K., \& Tsuruta, S. 1993, ApJ, 408, 186. 
\reference{}
Usov, V. V. 2002, astro-ph/0204402.
\reference{} 
Wang, F. Y.-H., \& Halpern, J. P. 1997, ApJ, 482, L159.
\reference{}
Xu, R. X., Qiao, G. J., \& Zhang, B. 1999, ApJ, 522, L109
\reference{} 
Zhang, L., \& Cheng, K. S. 1997, ApJ, 487, 370.
\reference{} 
Zhang, B. 2001, ApJ, 562, L59.
\reference{}
Zhang, B., Harding, A. K., \& Muslimov, A. G. 2000, ApJ, 531, L135.
\reference{} 
Zhang, B., \& Qiao, G. J. 1998, A \& A, 338, 62.
\end{references}
\end{document}